\documentclass[submission,copyright,creativecommons]{eptcs}
\providecommand{\event}{PROLE 2014} % Name of the event you are submitting to
\usepackage{breakurl}             % Not needed if you use pdflatex only.

\newtheorem{example}{Example}[section]
\usepackage[T1]{fontenc}

\usepackage[latin1]{inputenc}
\usepackage{graphicx}
\usepackage{flushend}
\normalsize
\usepackage{url}
\usepackage{latexsym}
\usepackage{amsfonts}
\usepackage{amssymb}
\usepackage[pdftex]{color}
\usepackage{colortbl}
\usepackage{listings}
\definecolor{light-gray}{gray}{0.95}  
\newcommand{\comment}[1]{}
\usepackage{dcolumn}
\newcolumntype{d}[1]{D{.}{\ldot}{#1}}
\newcolumntype{.}{D{.}{.}{-1}}
\usepackage{colortbl}
\newcommand*{\Scale}[2][4]{\scalebox{#1}{$#2$}}%

\title{XQOWL: An Extension of XQuery\\ for OWL Querying and Reasoning}

\author{Jes\'us M. Almendros-Jim\'enez\thanks{This work was supported by the EU (FEDER) and the Spanish MINECO Ministry ({\em Ministerio de Econom\'\i a y Competitividad}) under grant TIN2013-44742-C4-4-R,
as well as by the Andalusian Regional Government (Spain) under Project P10-TIC-6114.}
\institute{Dpto. de Inform\'atica\\
University of Almer\'\i a\\
04120-Almer\'\i a, SPAIN}
\email{jalmen@ual.es}
}
\def\titlerunning{XQOWL: An Extension of XQuery for OWL Querying and Reasoning
}
\def\authorrunning{Jes\'us M. Almendros-Jim\'enez}
\begin{document}
\maketitle

\begin{abstract}
One of the main aims of the so-called Web of Data is to be able to handle heterogeneous resources where 
data can be expressed in either XML or RDF. The design of programming languages able to handle both XML and RDF data
is a key target in this context. In this paper we present a framework called XQOWL that makes possible to handle XML and 
RDF/OWL data with XQuery. XQ\-OWL can be considered as an extension of the XQuery language
that connects XQuery with SPARQL and OWL reasoners. XQOWL embeds SPARQL queries (via Jena SPARQL engine) in XQuery and enables to make calls to OWL reasoners (HermiT, Pellet and FaCT++) from XQuery.
It permits to combine queries against XML and RDF/OWL resources
as well as to reason with RDF/OWL data. Therefore input data can be either XML or RDF/OWL
and output data can be formatted in XML (also using RDF/OWL XML serialization).
\end{abstract}

\section{Introduction}

There are two main formats to publish data on the Web. The first format is \emph{XML}, which is based on a tree-based model and for which the \emph{XPath} and \emph{XQuery} languages for querying, and the \emph{XSLT} language for transformation, have been proposed. The second format is \emph{RDF} which is a graph-based model and for which the \emph{SPARQL} language for querying and transformation has been proposed. Both formats (XML and RDF) can be used for describing data of a certain domain of interest. XML is used for instance in the \emph{Dublin Core} \footnote{\url{http://www.dublincore.org/.}}, \emph{MPEG-7} \footnote{\url{http://mpeg.chiariglione.org/.}}, among others, while RDF is used in \emph{DBPedia} \footnote{\url{http://www.dbpedia.org/.}}  and \emph{LinkedLifeData} \footnote{\url{http://linkedlifedata.com/.}}, among others. The number of organizations that offers their data from the Web is increasing in the last years. The so-called \emph{Linked open data} initiative\footnote{\url{http://linkeddata.org/}} aims to interconnect the published Web data. 

XML and RDF share the same end but they have different data models and query/transformation languages. Some data can be available in XML format and not in RDF format and vice versa. The \emph{W3C (World Wide Web Consortium)} \footnote{\url{http://www.w3.org/.}} proposes transformations from XML data to RDF data (called \emph{lifting}), and vice versa (called \emph{lowering}). RDF has XML-based representations (called \emph{serializations}) that makes possible to represent in XML the graph based structure of RDF.  However, XML-based languages are not usually used to query/transform serializations of RDF. Rather than SPARQL is used to query RDF whose syntax resembles SQL and abstract from the XML representation of RDF. The same happens when data are available in XML format: queries and transformations are usually expressed in XPath/XQuery/XSLT, instead of transforming XML to RDF, and using SPARQL.

One of the main aims of the so-called Web of Data is to be able to handle heterogeneous resources where 
data can be expressed in either XML or RDF. The design of programming languages able to handle both XML and RDF data
is a key target in this context and some recent proposals have been presented with this end. 
One of most known is XSPARQL \cite{XSPARQL} which is a hybrid language which
combines XQuery and SPARQL allowing to query XML and RDF. XSPARQL extends the XQuery syntax with 
new expressions able to traverse an RDF graph and construct the graph of the result of a query on RDF.
One of the uses of XSPARQL is the definition of lifting and lowering from XML to RDF and vice versa. 
But also XSPARQL is able to query XML and RDF data without transforming them, and obtaining the result
in any of the formats. They have defined a formal semantics for XSPARQL which is an extension of the XQuery semantics.
The SPARQL2XQuery interoperability framework \cite{SPARQL2XQuery} aims to overcome the same problem
by considering as query language SPARQL for both formats (XML and RDF), where SPARQL queries 
are transformed into XQuery queries by mapping XML Schemas into RDF metadata.
In early approaches, SPARQL queries are embedded in XQuery and XSLT \cite{Groppe} and XPath expressions
are embedded in SPARQL queries \cite{Droop}.

OWL is an ontology language working with concepts (i.e., classes) and roles (i.e., object/data properties) 
as well as with individuals (i.e., instances) which fill 
concepts and roles. OWL can be considered as an extension of RDF 
in which a richer vocabulary allows to express new relationships.
OWL offers more complex relationships than RDF between entities including means to limit the
properties of classes with respect to the number and type, means to infer that items with various
properties are members of a particular class, and a well-defined model of property inheritance.   
OWL reasoning \cite{handbook} is a topic of research of increasing interest in the literature. Most of OWL reasoners (for instance, {\it HermiT} \cite{HermiT}, {\it Racer} \cite{HaarslevRacer}, FaCT++ \cite{fact}, Pellet \cite{Pellet}) are based on tableaux based decision procedures. 

In this context, we can distinguish between (1) \emph{reasoning tasks} and (2) \emph{querying tasks} from a given ontology. The most typical (1) \emph{reasoning tasks}, with regard to a given ontology, include:
(a) \emph{instance checking}, that is, whether a particular individual is a member of a given concept, 
(b) \emph{relation checking}, that is,
whether two individuals hold a given role, 
(c) \emph{subsumption}, that is, whether a concept is a subset of another concept, 
(d) \emph{concept consistency}, that is, consistency of the concept relationships, and   
(e) a more general case of consistency checking is \emph{ontology consistency} in which the problem is to decide whether
a given ontology has a model. 
However, one can be also interested in (2) \emph{querying tasks} such as:
(a) {\it instance retrieval},  which means to retrieve all the individuals of a given concept, and
(b) {\it property fillers retrieval} which means to retrieve all the individuals which are related to a given individual 
with respect to a given role. 

SPARQL provides mechanisms for querying tasks while OWL reasoners are suitable for reasoning tasks.
SPARQL is a query language for RDF/OWL triples whose syntax resembles SQL.
OWL reasoners implement a complex deduction procedure including ontology consistency checking 
that SPARQL is not able to carry out. Therefore SPARQL/OWL reasoners are complementary 
in the world of OWL.

In this paper we present a framework called XQOWL that makes possible to handle XML and 
RDF/OWL data with XQuery. XQOWL can be considered as an extension of the XQuery language
that connects XQuery with SPARQL and OWL reasoners.
XQOWL embeds SPARQL queries (via Jena SPARQL engine) in XQuery
and enables to make calls to OWL reasoners (HermiT, Pellet and FaCT++) from XQuery.
It permits to combine queries against XML and RDF/OWL resources
as well as to reason with RDF/OWL data. Therefore input data can be either XML or RDF/OWL
and output data can be formatted in XML (also using RDF/OWL XML serialization).
We present two case studies: the first one which consists on lowering and lifting
similar to the presented in \cite{XSPARQL}; and the second one in which
XML analysis is carried out by mapping XML to an ontology and using a reasoner.
 
Thus the framework proposes to embed SPARQL code in XQuery as well as to make calls to OWL reasoners from XQuery.
With this aim a \emph{Java API} has been implemented on top of the OWL API \cite{owlapi}
and OWL Reasoner API \cite{OWLReasoner} that makes possible to interconnect XQuery with SPARQL
and OWL reasoners. The Java API is invoked from XQuery thanks to the use of the \emph{Java Binding}
facility available in most of XQuery processors (this is the case, for instance, of \emph{BaseX} \cite{BaseX}, \emph{Exist} 
\cite{Exist} and \emph{Saxon} \cite{Saxon}). The Java API enables
to connect XQuery to HermiT, Pellet and FaCT++ reasoners as well as to Jena SPARQL engine.
The Java API returns the results of querying and reasoning in XML format which can be handled
from XQuery.  It means that querying and reasoning RDF/OWL with XQOWL one can give XML format
to results in either XML or RDF/OWL. In particular, lifting and lowering is possible in XQOWL. 

Therefore our proposal can be seen as an extension of the proposed approaches for combining SPARQL
and XQuery. Our XQOWL framework is mainly focused on the use of XQuery for querying and reasoning
with OWL ontologies. It makes possible to write complex queries that combines SPARQL queries
with reasoning tasks. As far as we know our proposal is the first to provide such a combination.

The implementation has been tested with the BaseX processor \cite{BaseX} and can be downloaded
from our Web site \url{http://indalog.ual.es/XQOWL}. There the XQOWL API and the examples of
the paper are available as well as installation instructions.

Let us remark that here we continue our previous works on combination of XQuery and the Semantic Web.
In \cite{inap2009} we have described how to extend the syntax of XQuery in order to query RDF triples.
After, in \cite{apweb2011} we have presented a (Semantic Web) library for XQuery which makes possible to
retrieve the elements of an ontology as well as to use SWRL. Here, we have followed a new direction,
by embedding existent query languages (SPARQL) and reasoners in XQuery.

The structure of the paper is as follows. Section 2 will show an example of OWL ontology used in the
rest of the paper as running example. Section 3 will describe XQOWL: the Java API
as well as examples of use. Section 4 will present the case study of 
XML analysis by using an ontology.
Finally, Section 5 will conclude and present future work.

\section{OWL}\label{sec:owl}

{\scriptsize
\begin{table}[!t]
\begin{center}
\begin{tabular}{|l|l|l|}
 
\multicolumn{2}{|>{\columncolor[gray]{.8}}c|}{{\em Ontology}\quad }\\
(1) {\sf event}, {\sf message} $\sqsubseteq$ {\sf activity}  &
(2) {\sf wall}, {\sf album} $\sqsubseteq$ {\sf user\_item} \\
(3) {\sf $\forall$ created\_by.activity $\sqsubseteq$ user}   &
(4) {\sf added\_by}, {\sf sent\_by}  $\sqsubseteq$ {\sf created\_by}   \\
(5) {\sf $\top$ $\sqsubseteq$ $\leq$ 1. belongs\_to}   &
(6) {\sf $\forall$ belongs\_to.user\_item $\sqsubseteq$ user}  \\
(7) {\sf $\exists$ friend\_of.Self $\sqsubseteq$ $\bot$}    &
(8) {\sf friend\_of$^{-}$ $\sqsubseteq$ friend\_of}  \\
(9) {\sf $\forall$ friend\_of.user $\sqsubseteq$ user}   &
(10) {\sf $\forall$ invited\_to.user $\sqsubseteq$ event}   \\
(11) {\sf $\forall$ recommended\_friend\_of.user} & (12) {\sf friend\_of $\cdot$ friend\_of  $\sqsubseteq$}\\  {\sf \ \ \ \ $\sqsubseteq$ user}    &
{\sf \ \ \ \ recommended\_friend\_of }  \\
(13) {\sf $\exists$ replies\_to.Self $\sqsubseteq$ $\bot$}   &
(14) {\sf $\forall$ replies\_to.message $\sqsubseteq$ message}   \\
(15) {\sf $\top$ $\sqsubseteq$ $\leq$ 1.written\_in}   &
(16) {\sf $\forall$ written\_in.message $\sqsubseteq$ wall}    \\
(17) {\sf $\forall$ attends\_to.user $\sqsubseteq$ event}  &
(18) {\sf attends\_to$^{-}$ $\equiv$ confirmed\_by}   \\
(19) {\sf $\forall$ i\_like\_it.user $\sqsubseteq$ activity}   &
(20) {\sf i\_like\_it$^{-}$ $\equiv$ liked\_by}   \\
(21) {\sf $\forall$ content.message $\sqsubseteq$ String}   &
(22) {\sf $\forall$ date.event $\sqsubseteq$ DateTime}   \\
(23) {\sf $\forall$ name.event $\sqsubseteq$ String}  &
(24) {\sf $\forall$ nick.user $\sqsubseteq$ String}   \\
(25) {\sf $\forall$ password.user $\sqsubseteq$ String}  &
(26) ${\sf event \sqcap \exists confirmed\_by.user \sqsubseteq popular}$    \\
(27) ${\sf activity \sqcap}$ ${\sf \exists liked\_by.user \sqsubseteq popular}$   &
(28) $\sf activity \sqsubseteq {\leq 1}\  created\_by.user$    \\
(29) {\sf message $\sqcap$ event $\equiv$ $\bot$}   &\\
\hline
\end{tabular}
\caption{Social Network Ontology (in Description Logic Syntax)}\label{Ontology}
%\vspace*{-1cm}
\end{center}
\end{table}

}

In this section we show an example of ontology which will be used in the rest of the paper
as running example. 
Let us suppose an ontology about a social network (see Table \ref{Ontology}) in which
we define ontology classes: {\sf user}, {\sf user\_item}, {\sf activity}; and 
{\sf event}, {\sf message} $\sqsubseteq$ {\sf activity} (1); and {\sf wall}, {\sf album} $\sqsubseteq$ {\sf user\_item} (2).
In addition, we can define (object) properties as follows:
{\sf created\_by} which is a property whose domain is the class {\sf activity} and the range is {\sf user} (3),
and has two sub-properties:  {\sf added\_by}, {\sf sent\_by} (4)
(used for events and messages, respectively). 

We have also {\sf belongs\_to} which is a functional property (5) whose domain is {\sf user\_item} and range is {\sf user} (6);
{\sf friend\_of} which is a irreflexive (7) and symmetric (8) property whose domain and range is {\sf user} (9);
{\sf invited\_to} which is a property whose domain is {\sf user} and range is {\sf event} (10);
{\sf recommended\_friend\_of} which is a property whose domain and range is {\sf user} (11), and is the composition
of {\sf friend\_of} and {\sf friend\_of} (12);
{\sf replies\_to} which is an irreflexive property (13) whose domain and range is {\sf message} (14);
{\sf written\_in} which is a functional property (15) whose domain is {\sf message} and range is {\sf wall} (16);
{\sf attends\_to} which is a property whose domain is {\sf user} and range is {\sf event} (17) and is the inverse
of the property {\sf confirmed\_by} (18);
{\sf i\_like\_it} which is a property whose domain is {\sf user} and range is {\sf activity} (19), which is the inverse
of the property {\sf liked\_by} (20).

Besides, there are some (data) properties: the {\sf content} of a message (21), the {\sf date} (22) and {\sf name} (23)
of an event, and the {\sf nick} (24) and {\sf password} (25) of a user. Finally, we have defined the
concepts {\sf popular} which are events {\sf confirmed\_by} some user and activities
{\sf liked\_by} some user ((26) and (27)) 
and we have defined constraints: activities are {\sf created\_by} at most one user (28)
and {\sf message} and {\sf event} are disjoint classes (29).
Let us now suppose the set of individuals 
and object/data property instances of Table \ref{OntologyInstance}. 

{\scriptsize
\begin{table}[!t]
\begin{center}
\begin{tabular}{|l|l|l|}
 
\multicolumn{1}{|>{\columncolor[gray]{.8}}c|}{{\em Ontology Instance}\quad }\\
{\sf user}(jesus),  {\sf nick}(jesus,jalmen),\\
{\sf password}(jesus,passjesus), {\sf friend\_of}(jesus,luis)\\
\hline
{\sf user}(luis),  {\sf nick}(luis,Iamluis), {\sf password}(luis,luis0000)\\
\hline
{\sf user}(vicente),  {\sf nick}(vicente,vicente), {\sf password}(vicente,vicvicvic),\\ {\sf friend\_of}(vicente,luis),
  {\sf i\_like\_it}(vicente,message2),\\ {\sf invited\_to}(vicente,event1),
  {\sf attends\_to}(vicente,event1)\\
\hline
{\sf event}(event1),  {\sf added\_by}(event1,luis),\\ 
{\sf name}(event1,``Next conference''), {\sf date}(event1,21/10/2014)\\
\hline
{\sf event}(event2)\\
\hline
{\sf message}(message1),  {\sf sent\_by}(message1,jesus),\\
{\sf content}(message1,``I have sent the paper'')\\
\hline
{\sf message}(message2),  {\sf sent\_by}(message2,luis),\\
{\sf content}(message2,``good luck!''), {\sf replies\_to}(message2,message1)\\
\hline
{\sf wall}(wall\_jesus),  {\sf belongs\_to}(wall\_jesus,jesus)\\
\hline
{\sf wall}(wall\_luis),  {\sf belongs\_to}(wall\_luis,luis)\\
\hline
{\sf wall}(wall\_vicente),  {\sf belongs\_to}(wall\_vicente,vicente)\\
\hline
\end{tabular}
\caption{Individuals and object/data properties of the ontology}\label{OntologyInstance}
%\vspace*{-1cm}
\end{center}
\end{table}

}

From OWL reasoning we can deduce
new information. For instance, the individual
{\it message1} is an {\sf activity}, because {\sf message} is a subclass of {\sf activity},
and the individual {\it event1} is also an {\sf activity} because {\sf event} is a subclass of {\sf activity}. 
The individual {\it wall\_jesus} is an {\sf user\_item}
because {\sf wall} is a subclass of {\sf user\_item}. These inferences are obtained from the subclass
relation. In addition, object properties give us more information. For instance, the individuals {\it message1}, {\it message2} and
{\it event1} have been {\sf created\_by} {\it jesus}, {\it luis} and {\it luis}, respectively, since the properties
{\sf sent\_by} and {\sf added\_by}
are sub-properties of {\sf created\_by}. In addition, the individual {\it luis} is a {\sf friend\_of} {\it jesus} because {\sf friend\_of} is symmetric. 
More interesting is that the individual {\it vicente} is a {\sf recommended\_friend\_of}  {\it jesus}, because {\it jesus} is a {\sf friend\_of} {\it luis},
and {\it luis} is a {\sf friend\_of} {\it vicente}, which is deduced from the definition of {\sf recommended\_friend\_of},
which is the composition of {\sf friend\_of} and {\sf friend\_of}. 
Besides, the individual {\it event1} is {\sf confirmed\_by} {\it vicente}, because {\it vicente}
{\sf attends\_to} {\it event1} and the properties {\sf confirmed\_by} and {\sf attends\_to} are inverses. Finally, 
there are {\sf popular} concepts: {\it event1} and {\it message2}; the first one has been {\sf confirmed\_by} {\it vicente} and the second one is {\sf liked\_by} 
{\it vicente}.

The previous ontology is consistent. The ontology might introduce
elements that make the ontology inconsistent.  
We might add a user being {\sf friend\_of} of him(er) self. Even more, we can define that certain
events and messages are {\sf created\_by} (either {\sf added\_by} or {\sf sent\_by}) more than one user. Also a message
can reply to itself. However, there are elements that do not affect ontology consistency. For instance, {\it event2} has not
been {\sf created\_by} users. The ontology only requires to have at most one creator. Also, messages have not been
{\sf written\_in} a {\sf wall}. 

\section{XQOWL}

XQOWL allows to embed SPARQL queries in XQuery. It also makes possible to make calls to OWL reasoners.
With this aim a Java API has been developed.

\subsection{The Java API}

{\scriptsize
\begin{table}[!t]
\begin{center}
\begin{tabular}{|l|l|l|}
\multicolumn{1}{|>{\columncolor[gray]{.8}}c|}{{\em Java API}\quad }\\
\scriptsize public OWLReasoner getOWLReasonerHermiT(OWLOntology ontology)\\
\scriptsize public OWLReasoner getOWLReasonerPellet(OWLOntology ontology)\\
\scriptsize public OWLReasoner getOWLReasonerFact(OWLOntology ontology)\\
\scriptsize public String OWLSPARQL(String filei,String queryStr)\\
\scriptsize public <T extends OWLAxiom> String OWLQuerySetAxiom(Set<T> axioms)\\
\scriptsize public <T extends OWLEntity> String[] OWLQuerySetEntity(Set<T> elems)\\
\scriptsize public <T extends OWLEntity> String[] OWLReasonerNodeEntity(Node <T> elem)\\
\scriptsize public <T extends OWLEntity> String[] OWLReasonerNodeSetEntity(NodeSet<T> elems)\\
\hline
\end{tabular}
\caption{Java API of XQOWL}\label{Javalibrary}
%\vspace*{-1cm}
\end{center}
\end{table}

}

Now, we show the main elements of the Java API developed for connecting XQuery and SPARQL and
OWL reasoners. Basically, the Java API has been developed on top of the OWL API and the OWL Reasoner API 
and makes possible to retrieve results from SPARQL and OWL reasoners.
The elements of the library are shown in Table \ref{Javalibrary}. 

The first three elements of the library:
{\it getOWLReasonerHermiT}, {\it getOWLReasonerPellet} and  {\it getOWLReasonerFact}
make possible to instantiate HermiT,
Pellet and FaCT++ reasoners. For instance, the code of {\it getOWLReasonerHermiT} is as follows:

\begin{lstlisting}
public OWLReasoner getOWLReasonerHermiT(OWLOntology ontology){
		org.semanticweb.HermiT.Reasoner reasoner = new Reasoner(ontology); 
		reasoner.precomputeInferences(InferenceType.CLASS_HIERARCHY,
									InferenceType.CLASS_ASSERTIONS,
									...);
        return reasoner;	
	};
\end{lstlisting}

The fourth element of the library {\it OWLSPARQL} makes possible to instantiate
SPARQL Jena engine.  The input of this method is an ontology included in a file and a string representing
the SPARQL query. The output is a file (name) including the result of the query.
The code of {\it OWLSPARQL} is as follows:

\begin{lstlisting}
public String OWLSPARQL(String filei,String queryStr) 
throws FileNotFoundException{
	OntModel model = ModelFactory.createOntologyModel(); 
	model.read(filei);
	com.hp.hpl.jena.query.Query query = QueryFactory.create(queryStr);
	ResultSet result = 
			(ResultSet) SparqlDLExecutionFactory.create(query,model).execSelect();
	String fileName = "./tmp/"+result.hashCode()+"result.owl";
	File f = new File(fileName);
	FileOutputStream file = new FileOutputStream(f);
	ResultSetFormatter.outputAsXML(file,(com.hp.hpl.jena.query.ResultSet) result);
	try { file.close(); } catch (IOException e) {e.printStackTrace();}
		 return fileName;
	};
\end{lstlisting}

We can see in the code that the result of the query is obtained in XML format and stored in a file.
The rest of elements (i.e, {\it OWLQuerySetAxiom}, {\it  OWLQuerySetEntity}, {\it OWLReasonerNodeSetEntity}
and {\it OWLReasonerNodeEntity}) of the Java API make possible to handle the results
of calls to SPARQL and OWL reasoners. 
OWL Reasoners implement Java interfaces of the OWL API
for storing OWL elements. The main Java interfaces are {\it OWLAxiom} and {\it OWLEntity}.
{\it OWLAxiom} is a Java interface which is a super-interface of
all the types of OWL axioms:  {\it OWLSubClassOfAxiom}, {\it OWLSubDataPropertyOfAxiom}, {\it OWLSubObjectPropertyOfAxiom}, etc. {\it OWLEntity} is a Java interface which is a super-interface of all types
of OWL elements: {\it  OWLClass}, {\it OWLDataProperty}, {\it OWLDatatype}, etc. 

The XQOWL API includes the method {\it OWLQuerySetAxiom} 
that returns a file name where a set of axioms are included. 
It also includes {\it OWLQuerySetEntity} that returns in an array the URI's of a set of entities.  
Moreover, {\it OWLReasonerNodeEntity} returns in an array the URI's of a node.
Finally, {\it OWLReasonerNodeSetEntity} returns  in an array the URIs of a set of nodes.
For instance, the code of {\it OWLQuerySetEntity} is as follows:

\begin{lstlisting}
public <T extends OWLEntity> String[] OWLQuerySetEntity(Set<T> elems)    
		{	 
			String[] result = new String[elems.size()];
			Iterator<T> it = elems.iterator();
			for(int i=0;i<elems.size();i++){
				result[i]=it.next().toStringID();
							};
		return result;
		};
\end{lstlisting}

\subsection{XQOWL: SPARQL}

XQOWL is an extension of the XQuery language. Firstly, XQOWL allows
to write XQuery queries in which calls to SPARQL queries are achieved and the results of SPARQL
queries in XML format (see \cite{SPARQLXML}) can be handled by XQuery.
In XQOWL, XQuery variables can be bounded to results of SPARQL queries and vice versa,
XQuery bounded variables can be used in SPARQL expressions. Therefore, in XQOWL both XQuery and SPARQL
queries can share variables. 

\begin{example}
For instance, the following query returns the individuals
of concepts {\sf user} and {\sf event} in the social network:

\begin{lstlisting}
declare namespace spql="http://www.w3.org/2005/sparql-results#";
declare namespace xqo = "java:xqowl.XQOWL";

let $model := "socialnetwork.owl"
for $class in ("sn:user","sn:event")
return
let $queryStr := concat(
	"PREFIX rdf:  <http://www.w3.org/1999/02/22-rdf-syntax-ns#>
	 PREFIX sn: <http://www.semanticweb.org/socialnetwork.owl#>
	 SELECT ?Ind
	 WHERE { ?Ind rdf:type ", $class," }")
return
let $xqo := xqo:new()
let $res:= xqo:OWLSPARQL($xqo,$model,$queryStr)
return 
doc($res)/spql:sparql/spql:results/spql:result/spql:binding/spql:uri/text()
\end{lstlisting}

Let us observe that the name of the classes (i.e., {\sf sn:user} and {\sf sn:event}) 
is defined by an XQuery variable (i.e., {\tt \$class}) in a {\tt for} expression, which is
passed as parameter of the SPARQL expression. In addition, the result is obtained in an XQuery variable
(i.e. {\tt \$res}). Here {\it OWLSPARQL} of the XQOWL API is used to call the SPARQL Jena engine, which returns a file name
(a temporal file) in which the result is found. Now, {\tt \$res} can be used from XQuery to obtain the URIs of the elements:
$$\Scale[0.80] {\tt doc(\$res)/spql:sparql/spql:results/spql:result/spql:bin\-ding\-/spql:uri/text()}$$
In this case, we obtain the following plain text:

\begin{lstlisting}
http://www.semanticweb.org/socialnetwork.owl#vicente
http://www.semanticweb.org/socialnetwork.owl#jesus
http://www.semanticweb.org/socialnetwork.owl#luis
http://www.semanticweb.org/socialnetwork.owl#event2
http://www.semanticweb.org/socialnetwork.owl#event1
\end{lstlisting}
\end{example}

\begin{example}
Another example of using XQOWL and SPARQL is the code of {\it lowering} from the document:

\begin{lstlisting}
<rdf:RDF xmlns:rdf="http://www.w3.org/1999/02/22-rdf-syntax-ns#" 
	xmlns="http://relations.org">
  <foaf:Person xmlns:foaf="http://xmlns.com/foaf/0.1/" rdf:about="#b1">
    <foaf:name>Alice</foaf:name>
    <foaf:knows>
      <foaf:Person rdf:about="#b4"/>
    </foaf:knows>
    <foaf:knows>
      <foaf:Person rdf:about="#b6"/>
    </foaf:knows>
  </foaf:Person>
  <foaf:Person xmlns:foaf="http://xmlns.com/foaf/0.1/" rdf:about="#b4">
    <foaf:name>Bob</foaf:name>
    <foaf:knows>
      <foaf:Person rdf:about="#b6"/>
    </foaf:knows>
  </foaf:Person>
  <foaf:Person xmlns:foaf="http://xmlns.com/foaf/0.1/" rdf:about="#b6">
    <foaf:name>Charles</foaf:name>
  </foaf:Person>
</rdf:RDF>
\end{lstlisting}

\noindent to the document: 

\begin{lstlisting}
<relations>
<person name="Alice">
<knows> Bob </knows>
<knows> Charles </knows>
</person>
<person name="Bob">
<knows> Charles </knows>
</person>
<person name="Charles" />
</relations>
\end{lstlisting}

\noindent This example has been taken from \cite{XSPARQL}\footnote{XSPARQL works with blank nodes, and there the RDF document
includes {\tt nodeID} tag for each RDF item. In XQOWL we cannot deal with blank nodes at all, and therefore
a preprocessing of the RDF document is required: {\tt nodeID} tags are replaced by {\tt about}.} in which
they show the lowering example in XSPARQL. In our case the code
of the lowering example is as follows:

\begin{lstlisting}
declare namespace spql="http://www.w3.org/2005/sparql-results#";
declare namespace xqo = "java:xqowl.XQOWL";
declare variable $model := "relations.rdf";

let $query1 :=  
		"PREFIX rdfs: <http://www.w3.org/2000/01/rdf-schema#> 
		PREFIX rdf:  <http://www.w3.org/1999/02/22-rdf-syntax-ns#>
		PREFIX foaf: <http://xmlns.com/foaf/0.1/>
		SELECT ?Person ?Name  
		WHERE {
		?Person foaf:name ?Name
		} ORDER BY ?Name"
let $xqo := xqo:new(),
$result := xqo:OWLSPARQL($xqo,$model,$query1)
return  
for $Binding in doc($result)/spql:sparql/spql:results/spql:result
let $Name := $Binding/spql:binding[@name="Name"]/spql:literal/text(),
    $Person := $Binding/spql:binding[@name="Person"]/spql:uri/text(),
    $PersonName := functx:fragment-from-uri($Person)
return
<person name="{$Name}">{
let  $query2 :=
	concat(
	"PREFIX rdfs: <http://www.w3.org/2000/01/rdf-schema#>
	PREFIX rdf:  <http://www.w3.org/1999/02/22-rdf-syntax-ns#>
	PREFIX rel:  <http://relations.org#>
	PREFIX foaf: <http://xmlns.com/foaf/0.1/>
	SELECT ?FName
	WHERE { 
              _:",$PersonName," foaf:knows ?Friend .  
              _:",$PersonName," foaf:name ", "'",$Name,"' .  
              ?Friend foaf:name ?FName 
}")
let $result2 := xqo:OWLSPARQL($xqo,$model,$query2)
return  
for $FName in doc($result2)/spql:sparql/spql:results/spql:result/spql:binding/spql:literal/text()
return
<knows>{$FName}</knows>
}
</person>
}
</relations>
\end{lstlisting}
In this example, two SPARQL queries are nested and share variables. The result of the first SPARQL query
(i.e., {\tt \$PersonName} and {\tt \$Name}) is used in the second SPARQL query. 
\end{example}

\subsection{XQOWL: OWL Reasoners}

XQOWL can be also used for querying and reasoning with OWL. With this aim the OWL API and OWL Reasoner
API have been integrated in XQuery. Also for this integration, the XQOWL API is required. For using OWL
Reasoners from XQOWL there are some calls to be made from XQuery code. Firstly, we have to instantiate 
the ontology manager by using {\it createOWLOntologyManager}; secondly, the ontology has to be loaded
by using {\it loadOntologyFromOntologyDocument}; thirdly, in order to handle OWL elements we have to
instantiate the data factory by using {\it getOWLDataFactory}; finally, in order to select a reasoner
{\it getOWLReasonerHermiT}, {\it getOWLReasonerPellet} and  {\it getOWLReasonerFact} are used.

\begin{example}
For instance, we can query the object properties of the ontology using the OWL API as follows:

%declare namespace rdf="http://www.w3.org/1999/02/22-rdf-syntax-ns#";
%declare namespace owl="http://www.w3.org/2002/07/owl#";
%declare namespace om = 
%			"java:org.semanticweb.owlapi.model.OWLOntologyManager";
%declare namespace o = "java:org.semanticweb.owlapi.model.OWLOntology";
%declare namespace xqo = "java:xqowl.XQOWL";
%declare namespace file = "java:java.io.File";
%declare namespace api = "java:org.semanticweb.owlapi.apibinding.OWLManager";
%declare namespace sn = "http://www.semanticweb.org/ontologies/2011/7/socialnetwork.owl#";
%declare variable $file := "socialnetwork.owl";

\begin{lstlisting}
let $xqo := xqo:new(),
		$man := api:createOWLOntologyManager(),
		$fileName := file:new($file),
		$ont := om:loadOntologyFromOntologyDocument($man,$fileName)
return 
doc(xqo:OWLQuerySetAxiom($xqo,o:getAxioms($ont)))/rdf:RDF/owl:ObjectProperty 
\end{lstlisting}
\noindent obtaining the following result:
\begin{lstlisting}
<ObjectProperty...rdf:about="...#added_by">
	<rdfs:subPropertyOf rdf:resource="...#created_by"/>
	<rdfs:domain rdf:resource="...#event"/>
	<rdfs:range rdf:resource="...#user"/>
</ObjectProperty>
<ObjectProperty ... rdf:about="...#attends_to">
	<inverseOf rdf:resource="...#confirmed_by"/>
	<rdfs:range rdf:resource="...#event"/>
	<rdfs:domain rdf:resource="...#user"/>
</ObjectProperty>
...
\end{lstlisting}

\end{example}

\begin{example}
Another example of query using the OWL API is the following which requests
class axioms related to {\sf wall} and {\sf event}:

%declare namespace owl="http://www.w3.org/2002/07/owl#";
%declare namespace rdf="http://www.w3.org/1999/02/22-rdf-syntax-ns#";
%declare namespace om = 
%		"java:org.semanticweb.owlapi.model.OWLOntologyManager";
%declare namespace o = 
%		"java:org.semanticweb.owlapi.model.OWLOntology";
%declare namespace xqo = "java:xqowl.XQOWL";
%declare namespace df = 
%		"java:org.semanticweb.owlapi.model.OWLDataFactory";
%declare namespace file = "java:java.io.File";
%declare namespace iri = "java:org.semanticweb.owlapi.model.IRI";
%declare namespace api = 
%		"java:org.semanticweb.owlapi.apibinding.OWLManager";
%declare variable $base := 
%		"http://www.semanticweb.org/ontologies/2011/7/socialnetwork.owl#";
%declare variable $file := "socialnetwork.owl";

\begin{lstlisting}
let $xqo := xqo:new(),
	$man := api:createOWLOntologyManager(),
	$fileName := file:new($file),
	$ont := om:loadOntologyFromOntologyDocument($man,$fileName),
	$fact := om:getOWLDataFactory($man)
return
for $class in ("wall","event")
let $iri := iri:create(concat($base,$class)),
	$class := df:getOWLClass($fact,$iri)
return 
doc(xqo:OWLQuerySetAxiom($xqo,o:getAxioms($ont,$class)))/rdf:RDF/owl:Class  
\end{lstlisting}
\noindent in which a {\tt for} expression is used to define the names of the classes to
be retrieved, obtaining the following result:
\begin{lstlisting}
<Class ... rdf:about="...#user_item"/>
<Class ... rdf:about="...#wall">
  <rdfs:subClassOf rdf:resource="...#user_item"/>
</Class>
<Class ... rdf:about="...#activity"/>
<Class ... rdf:about="...#event">
  <rdfs:subClassOf rdf:resource="...#activity"/>
  <disjointWith rdf:resource="...#message"/>
</Class>
<Class ... rdf:about="...#message"/>
\end{lstlisting}
\end{example}

Now we can see examples about how to use XQOWL for reasoning with an ontology. With this aim,
we can use the OWL Reasoner API (as well as the XQOWL API).  The XQOWL API allows easily 
to use HermiT, Pellet and FaCT++ reasoners. 

\begin{example}
For instance, let us suppose we want to check
the consistence of the ontology by the HermiT reasoner. The code is as follows:

\begin{lstlisting}
let $xqo := xqo:new(),
	$man := api:createOWLOntologyManager(),
	$fileName := file:new($file),
	$ont := om:loadOntologyFromOntologyDocument($man,$fileName),
	$fact := om:getOWLDataFactory($man),
	$reasoner := xqo:getOWLReasonerHermiT($xqo,$ont),
	$boolean := r:isConsistent($reasoner),
	$dispose := r:dispose($reasoner)
return $boolean
\end{lstlisting}

\noindent which returns {\tt true}. Here the HermiT reasoner is instantiated by using {\it getOWLReasonerHermiT}.
In addition, the OWL Reasoner API method {\it isConsistent} is used to check ontology consistence. Each
time the work of the reasoner is done, a call to {\it dispose} is required.
\end{example}

\begin{example}
Let us suppose now we want to retrieve instances of concepts {\tt activity} and {\tt user}. Now, 
we can write the following query using the HermiT reasoner:

\begin{lstlisting}
for $classes in ("activity","user")
let $xqo := xqo:new(),
	$man := api:createOWLOntologyManager(),
	$fileName := file:new($file),
	$ont := om:loadOntologyFromOntologyDocument($man,$fileName),
	$fact := om:getOWLDataFactory($man),
	$iri := iri:create(concat($base,$classes)),
	$reasoner := xqo:getOWLReasonerHermiT($xqo,$ont),
	$class := df:getOWLClass($fact,$iri),
	$result:= r:getInstances($reasoner,$class,false()),
	$dispose := r:dispose($reasoner)
return
<concept class="{$classes}">
{ for $instances in xqo:OWLReasonerNodeSetEntity($xqo,$result) 
    return <instance>{substring-after($instances,'#')}</instance>}
</concept>
\end{lstlisting}
\noindent obtaining the following result in XML format:
\begin{lstlisting}
<concept class="activity">
  <instance>message1</instance>
  <instance>message2</instance>
  <instance>event1</instance>
  <instance>event2</instance>
</concept>
<concept class="user">
  <instance>jesus</instance>
  <instance>vicente</instance>
  <instance>luis</instance>
</concept>
\end{lstlisting}
Here {\it getInstances} of the OWL Reasoner API is used to retrieve the instances of a given ontology class.
In addition, a call to {\it create} of the OWL API, which creates the IRI of the class,
and a call to {\it getClass} of the OWL API, which retrieves the class, are required. The OWL Reasoner
is able to deduce that {\tt message1} and {\tt message2} belong to concept {\tt activity} since
they belong to concept {\tt message} and {\tt message} is a subconcept of {\tt activity}. The same
can be said for events. 
\end{example}

\begin{example}
Let us suppose now we want to retrieve the subconcepts of {\tt activity} using the Pellet reasoner. The code
is as follows:

\begin{lstlisting}
let $xqo := xqo:new(),
	$man := api:createOWLOntologyManager(),
	$fileName := file:new($file),
	$ont := om:loadOntologyFromOntologyDocument($man,$fileName),
	$fact := om:getOWLDataFactory($man),
	$iri := iri:create(concat($base,"activity")),
	$reasoner := xqo:getOWLReasonerPellet($xqo,$ont),
	$class := df:getOWLClass($fact,$iri),
	$result:= r:getSubClasses($reasoner,$class,false()),
	$dispose := r:dispose($reasoner)
return 
	for $subclass in xqo:OWLReasonerNodeSetEntity($xqo,$result) 
		return <subclass>{substring-after($subclass,'#')} </subclass>
\end{lstlisting} 

\noindent and the result in XML format is as follows:
\begin{lstlisting}
<subclass>popular_message</subclass>
<subclass>event</subclass>
<subclass>Nothing</subclass>
<subclass>popular_event</subclass>
<subclass>message</subclass>
\end{lstlisting} 
Here {\it getSubClasses} of the OWL Reasoner API is used.
\end{example}

\begin{example}
Finally, let us suppose we want to retrieve the recommended friends of {\tt jesus}. Now, the query is as follows:

\begin{lstlisting}
let $xqo := xqo:new(),
	$man := api:createOWLOntologyManager(),
	$fileName := file:new($file),
	$ont := om:loadOntologyFromOntologyDocument($man,$fileName),
	$fact := om:getOWLDataFactory($man),
	$iri := iri:create(concat($base,"recommended_friend_of")),
	$iri2 := iri:create(concat($base,"jesus")),
	$reasoner := xqo:getOWLReasonerPellet($xqo,$ont),
	$property := df:getOWLObjectProperty($fact,$iri),
	$ind := df:getOWLNamedIndividual($fact,$iri2),
	$result:= r:getObjectPropertyValues($reasoner,$ind,$property),
	$dispose := r:dispose($reasoner)
return 
for $rfriend in xqo:OWLReasonerNodeSetEntity($xqo,$result)
return
<recommended_friend>
{substring-after($rfriend,'#')}
</recommended_friend>
\end{lstlisting}
\noindent and the answer as follows:
\begin{lstlisting}
<recommended_friend>jesus</recommended_friend>
<recommended_friend>vicente</recommended_friend>
\end{lstlisting}
Here the OWL Reasoner API is used to deduce the friends of friends of {\tt jesus}. Due to symmetry of {\tt friend}
relationship, a person is a recommended friend of itself.
\end{example}

\section{Using XQOWL for XML Analysis}

Now, we show an example in which XQOWL is used to analyze the semantic content of an XML document.
This example was used in our previous work \cite{odbase2012} to illustrate the use of our Semantic Web library 
for XQuery.  The example takes an XML document as input as follows:

%\begin{figure*}[!t]
\begin{lstlisting}[morekeywords={conference,papers,title,name,researchers,paper,researcher,wordCount,studentPaper,isStudent,manuscript,referee}]
<?xml version='1.0'?>
<conference>
<papers>
<paper id="1" studentPaper="true">
<title> XML Schemas </title>
<wordCount> 1200 </wordCount>
</paper>
<paper id="2"  studentPaper="false">
<title> XML and OWL </title>
<wordCount> 2800 </wordCount>
</paper>
<paper id="3" studentPaper="true">
<title> OWL and RDF </title>
<wordCount> 12000 </wordCount>
</paper>
</papers>
<researchers>
<researcher id="a" isStudent="false" manuscript="1" referee="1">
<name>Smith </name>
</researcher>
<researcher id="b" isStudent="true" manuscript="1" referee="2">
<name>Douglas </name>
</researcher>
<researcher id="c" isStudent="false" manuscript="2" referee="3">
<name>King </name>
</researcher>
<researcher id="d" isStudent="true" manuscript="2" referee="1">
<name>Ben </name>
</researcher>
<researcher id="e" isStudent="false" manuscript="3" referee="3">
<name>William</name>
</researcher>
</researchers>
</conference>
\end{lstlisting}
%\caption{Running example \label{papers}}
%\end{figure*}

The document lists {\it paper}s and {\it researcher}s
involved in a {\it conference}. Each {\it paper} and {\it researcher} has an identifier (represented by the attribute {\it id}),
and has an associated set of labels: {\it title} and {\it wordCount} for {\it paper}s and {\it name} for {\it researcher}s.
Furthermore, they have attributes {\it studentPaper} for {\it paper}s and {\it isStudent}, {\it manuscript} and {\it referee}
for {\it researcher}s.  The meaning of {\it manuscript} and {\it referee} is that the given researcher has submitted
the paper of number described by {\it manuscript} as well as has participated as reviewer of the paper of number
given by {\it referee}. 

Now, let us suppose that we would like to analyze the content of the XML document
in order to detect constraints which are violated. 
In particular, the revision system of the conference forbids that an student is a reviewer as well as a research is
a reviewer of his(her) own paper.  

In order to analyze the document the idea is to create an ontology 
to represent the same elements of the XML document. This ontology contains in the {\bf TBox} a vocabulary
to represent submissions. It includes class names {\it Paper} and {\it Researcher}. But also it includes {\it PaperofSenior},
{\it PaperofStudent}, {\it Student} and {\it Senior}. The individuals of {\it PaperofSenior} are the papers for which
{\it studentPaper} of the XML document has been set to false. The individuals of {\it PaperofStudent} are the papers for which
{\it studentPaper} of the XML document has been set to true. Analogously, the individuals of {\it Senior}
and {\it Student} are the researchers for which {\it isStudent} has been set to false, respectively, to true.
In addition the ontology includes
object properties {\it manuscript} and {\it referee}, and data properties {\it wordCount}, {\it name} and {\it title}.

Now, the idea is to express the revision system constraints as constraints of the ontology.
Thus, the ontology includes two restrictions to be checked: {\it Student} and {\it Reviewer} classes are disjoint
while {\it manuscript} and {\it referee} are disjoint object properties.

In order to analyze a given XML document, we can use XQOWL with two ends.

\begin{itemize}
%\item To create the ontology {\bf TBox}.
\item To transform the XML document to the ontology {\bf ABox}.
\item To check consistence of the ontology.
\end{itemize} 

\noindent The code of the transformation to the ontology {\bf ABox} is as follows: 

\begin{lstlisting}
let $name := /conference 
let $ontology1 :=
	(for $x in $name/papers/paper return 
		sw:toClassFiller(sw:ID($x/@id),"#Paper") union 
		(
		let $studentPaper:= $x/@studentPaper return 
			if (data($studentPaper)="true") then
				sw:toClassFiller(sw:ID($x/@id),"#PaperofStudent")
				else sw:toClassFiller(sw:ID($x/@id),"#PaperofSenior")
		) union 
		sw:toDataFiller(sw:ID($x/@id),"title",$x/title,"string") union 
		sw:toDataFiller(sw:ID($x/@id),"wordCount",$x/wordCount,"integer")
) 
let $ontology2 :=
(for $y in $name/researchers/researcher return 
		sw:toClassFiller(sw:ID($y/@id),"#Researcher") union 
		sw:toDataFiller(sw:ID($y/@id),"name",$y/name,"string") union
		(
		let $student:= $y/@isStudent return 
			if (data($student)="true") then
				sw:toClassFiller(sw:ID($y/@id),"#Student")
				else sw:toClassFiller(sw:ID($y/@id),"#Senior")
		) union
		sw:toObjectFiller(sw:ID($y/@id),"manuscript",sw:ID($y/@manuscript)) union 
		sw:toObjectFiller(sw:ID($y/@id),"referee",sw:ID($y/@referee)))
return   
let $mapping :=  $ontology1 union $ontology2
return 
let $doc :=
document{
<rdf:RDF ...>
	{doc("ontology_papers.owl")/rdf:RDF/*}
	{$mapping}
</rdf:RDF>
}
\end{lstlisting}

Here we have used the Semantic Web library for XQuery defined in \cite{odbase2012}.
Basically, we have created the instance of the ontology by using {\it sw:toClassFiller},
{\it sw:toDataFiller} and {\it sw:toObjectFiller} which make possible to create instances
of classes, data and object properties, respectively. At the end of the code,
the ontology {\bf TBox} is incorporated (which is stored in the file {\it ``ontology\_papers.owl''}).
Now, the consistence checking using the Hermit reasoner is as follows, where {\tt \$doc} is the result of the previous query:

\begin{lstlisting}
let $xqo := xqo:new(),
$man := api:createOWLOntologyManager(),
$seq := file:write("ontology_analysis.owl",$doc),
$fileName := file_io:new($file),
$ont := om:loadOntologyFromOntologyDocument($man,$fileName),
$fact := om:getOWLDataFactory($man),
$reasoner := xqo:getOWLReasonerHermit($xqo,$ont),
$boolean := r:isConsistent($reasoner),
$dispose := r:dispose($reasoner)
return $boolean
\end{lstlisting}

\section{Conclusions and Future Work}\label{sec:conclusions}

In this paper we have presented an extension of XQuery called XQOWL to query XML and RDF/OWL documents, as well as
to reason with RDF/OWL resources. We have described the XQOWL API that allows to make calls from XQuery 
to SPARQL and OWL Reasoners. Also we have shown examples of use of XQOWL. The main advantage
of the approach is to be able to handle both types of documents through the sharing of variables between 
XQuery and SPARQL/OWL Reasoners. 
The implementation has been tested with the BaseX processor \cite{BaseX} and can be downloaded
from our Web site \url{http://indalog.ual.es/XQOWL}. 
As future work, we would like to extend our work as follows.
Firstly, we would like to extend our Java API. More concretely, with the SWRL API in order to execute rules
from XQuery, and to be able to provide explanations about ontology inconsistence. Secondly, we would like
to use our framework in ontology transformations (refactoring, coercion, splitting, amalgamation) and matching.

\bibliographystyle{eptcs}
%\bibliography{biblio}
\bibliography{references}
\end{document}